# High yield fabrication of chemically reduced graphene oxide field effect transistors by dielectrophoresis

Daeha Joung[1,2], A. Chunder[1,3], Lei Zhai[1,3], and Saiful I. Khondaker[1,2] *

[1] Nanoscience Technology Center, [2] Department of Physics, [3] Department of Chemistry, University of Central Florida, Orlando, Florida 32826, USA.

* To whom correspondence should be addressed. E-mail: saiful@mail.ucf.edu

**Abstract**

We demonstrate high yield fabrication of field effect transistors (FET) using chemically reduced graphene oxide (RGO) sheets suspended in water assembled via dielectrophoresis. The two terminal resistances of the devices were improved by an order of magnitude upon mild annealing at 200 $^0$C in Ar/H$_2$ environment for 1 hour. With the application of a backgate voltage, all of the devices showed FET behavior with maximum hole and electron mobilities of 4.0 and 1.5 cm$^2$/Vs respectively. This study shows promise for scaled up fabrication of graphene based nanoelectronic devices.

Graphene has attracted a great deal of attention because of its unique electronic properties making them model systems for the observation of novel quantum phenomenon and building blocks for future nano electronic devices [1-4]. The majority of the existing graphene based electronic devices involve a mechanical exfoliation technique by peeling layers from highly oriented pyrolitic graphite (HOPG) using scotch tape. After deposition onto SiO$_2$ substrates, suitable graphene sheets were located using an optical microscope or atomic force microscope (AFM) followed by making electrical contact. Although this technique offers high quality devices for studying novel physics, however, the extremely low throughput and lack of methods for precise positioning of the sheets severely limits their large scale integration into devices. For practical applications in nanoelectronic devices, it is necessary to (i) create graphene nanostructures in large quantities and (ii) integrate them at selected positions of the circuits with high yield. Solution processing of graphene provides an attractive approach to produce graphene sheets in large quantities at low cost [5-12]. The easy processability and compatibility with various substrate including plastics also makes them attractive candidate for high yield manufacturing of graphene based circuits.

Although production of graphene in large quantity can be achieved via solution processing technique, however, the assembly of graphene devices at selected position of the circuit with high yield is still at its infancy. There are only a few reports exist on the directed assembly of graphene nanostructures from solution which only concentrated on the assembly and not on functional devices. Burg et. al. [13] demonstrated dielectrophoretic (DEP) assembly of few layers (3-10) of insulating GO which became metallic upon thermal reduction at 450 $^0$C. Kang et. al. [14] carried out DEP assembly of 4 - 13 layers of GO sheets and reduced the GO devices either chemically using hydrazine or thermally at up to 1000 $^0$C to restore electrical conductivity. Vijayaraghavan et. al. [15] also demonstrated DEP assembly of few layers graphene flakes and presented current-voltage characteristics. Field effect transistor (FET) behavior has not been observed in any of these studies which are critical for the fabrication of functional electronic circuits. An alternative approach for high yield fabrication of graphene devices could be based upon chemical reduction of GO sheets in the solution and then dielectrophoretically assemble them into devices. Such an approach could be advantageous and compatible with the current Complimentary Metals Oxide Semiconductor (CMOS) technology as it would not require high temperature post-assembly reduction to recover the electrical conductivity of the GO sheets.

In this paper, we report on the high yield CMOS compatible FETs fabrication using chemically reduced grapheme oxide (RGO) sheets from solution via ac dielectrophoresis (DEP). The GO sheets suspended in water were reduced chemically in the solution phase and then assembled between prefabricated gold source and drain electrodes by applying an ac voltage of 3 $V_{p-p}$ with a frequency of 1 MHz. Atomic force microscopy investigation after assembly showed that a few layers (5-15 layers) of graphene were assembled with 100% device yield. The two terminal resistances of the devices improved by almost an order of magnitude upon mild annealing at 200 $^0$C in argon/hydrogen (Ar/H$_2$) environment for 1 hour. With the application of a backgate voltage, all of the devices showed FET behavior with maximum hole and electron mobilities of 4.0 and 1.5 cm$^2$/Vs respectively. This study is a significant step forward in scaled up fabrication of graphene based nanoelectronic devices.

RGO sheets were synthesized through a reduction of GO prepared by modified Hummers method [16]. Oxidized graphite in water was ultrasonicated to achieve GO sheets followed by centrifugation for 30 minutes at 3000 rpm to remove any unexfoliated oxidized graphite. The pH of GO dispersion in water (0.1 mg/ml) was adjusted to 11 using a 5% ammonia aqueous solution. 15 μl of hydrazine solution (35% in DMF) was then added to the mixture. The mixture was heated at 95-100 $^0$C for 1 hour and cooled to room temperature. X-ray photoelectron spectroscopy (XPS) spectra taken before and after the reduction show that the intensity ratios of the C-C and C-O bonds changed from 1.4 to 8.0 verifying effective reduction of GO (see supporting information for XPS results of GO and RGO) [17] and are consistent with previous observations [12,18,19]. The RGO suspension was spin coated on a mica substrate and examined using AFM. Figure 1(a) displays a tapping-mode AFM image of the RGO sheets along with their height analysis. The lateral dimension of our RGO sheets varies from 0.2 -1 μm. The line graph represents the thickness of the RGO sheets. Approximately 70 % of the sheets displayed a height of 1.0 ±0.2 nm. This height is similar to other single layer RGO sheet height reported in previous studies [6-8].

Devices were fabricated on heavily doped silicon substrates capped with a thermally grown 250 nm thick SiO$_2$ layer. The electrode patterns were fabricated by a combination of optical and electron beam lithography (EBL). First, contact pads with larger patterns and electron beam markers were fabricated with optical lithography using double layer resists (LOR 3A/Shipley 1813) developing in CD26, followed by thermal evaporation of chromium (Cr) (5 nm) and Au (45 nm) and finally standard lift-off. Smaller patterns were then defined with EBL using single layer PMMA resists and then developed in (1:3) methyl isobutyl ketone : isopropal alchohol (MIBK:IPA). After defining the patterns, 5 nm thick Cr and 20 nm thick Au were thermally deposited followed by lift-off. The channel lengths and widths were 500 nm x 500 nm. Prior to the DEP assembly, the electrodes were treated in oxygen plasma for 15 minutes to remove any residual organics.

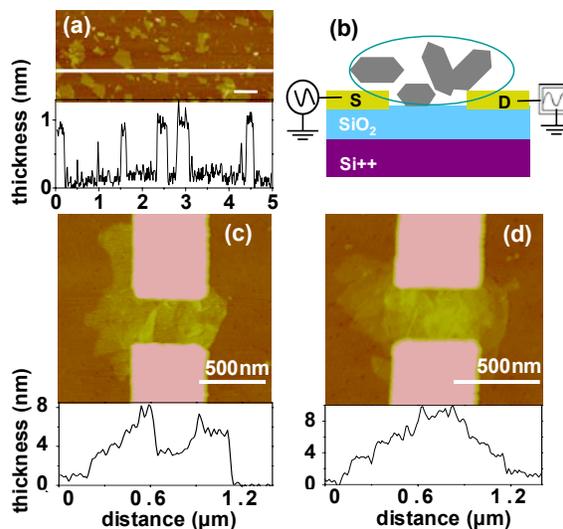

FIG. 1. (Color online) (a) Tapping-mode AFM images of RGO sheets with a height profile indicating majority of the sheets are single layer. (b) Cartoon of DEP assembly set-up. (c)-(d) Tapping-mode AFM of two RGO device assembled via DEP along with their height profile. The thickness varies from 2 nm to 10 nm in the channel indicating that up to 10 layers of RGO sheets have been assembled in the channel. Scale bar represents 500 nm.

The DEP assembly of RGO sheets was carried out in a probe station under ambient condition. Figure 1 (b) shows a cartoon of the DEP set up. A small drop of RGO (2 μl) solution was placed onto a chip containing 18 pairs of source and drain electrodes. An AC voltage of approximately 3 $V_{P-P}$ at 1 MHz was applied with a function generator for 20-30 seconds to the electrode pair and then moved to the next pair. The AC voltage gave rise to a time averaged DEP force given by $F_{DEP} = (p \cdot \nabla)E$, where p is the induced dipole moment of the polarizable object and E is the non-uniform electric field [20]. The strong electric field gradient caused the RGO sheets to align along the field direction and assembled between the source and drain electrodes. After the assembly, the remaining solution was blown off by nitrogen gas and the chips were analyzed by atomic force microscopy (AFM). Figure 1(c) and 1(d) show tapping-mode AFM image of two representative devices along with their height analysis. It can be seen that the thickness varies from 2 nm to 10 nm in the channel indicating that up to 10 layers of RGO sheets have been assembled in the channel. The thickness is lower at the edges, demonstrating one or two layers of graphene sheet near the edge while the thickness is higher in the middle of the channel due to overlap of several individual sheets or folding of sheets. Among the 100 pairs of electrodes that we have used for the DEP assembly, we found that all of the electrodes were bridged by a few layers of RGO sheets giving a 100% device yield (see supporting information for the electrode geometry and scanning electron microscope (SEM) image of the devices from one chip) [17]. The maximum thickness of RGO sheets in the channel varied between 5 nm to 15 nm from device to device. It is not clear why few layers of RGO sheets have been assembled in all the devices despite a large number of single layer RGO is present in the solution. Currently we are experimenting with different electrode geometries, trapping times, voltages and frequencies to see whether single layer RGO sheets can be assembled using DEP. We note that previous DEP studies of GO sheets also found multiple layers of GO sheets assembled during the assembly [13-14].

Following the DEP assembly, the room temperature electronic properties of the RGO devices were carried out in a probe station. Figure 2 (a) shows a sketch of the electrical transport measurement set-up. The measurements were performed using a Keithley 2400 source-meter, and a current preamplifier (DL 1211) capable of measuring sub-pA signal interfaced with LABVIEW program. The highly doped silicon was used as a global backgate. Figure 2 (b) shows a representative plot of drain current (I) versus source-drain voltage (V) for one of the RGO devices. The dotted curve represents the as-assembled device while the solid curve represents after mild thermal annealing. The thermal annealing was done in Ar/H$_2$ gas at 200 $^0$C for an hour. From the linear I-V curves a resistance of 10 MΩ and 980 kΩ was calculated for before and after mild annealing, respectively, with improvement by a factor of 10. All of our devices show similar improvement upon annealing. The measured conductivities of our RGO devices were about 200 S/m.

Figure 2 (c) and 2 (d) show transfer characteristics of two representative RGO FETs (device A and device B) where current ($I$) is plotted as a function of gate voltage ($V_g$) with fixed source-drain voltage $V = 1$ V. The field-effect mobility, $\mu$, was estimated from the $I$ – $V_g$ relation as: $\mu = (L/WC_{ox}V)/(\Delta I/\Delta V_g)$ where, $L$ is channel length, $W$ is channel width and $C_{ox}$ is capacitance per unit area of the gate insulator. Figure 2 (c) shows the $I$-$V_g$ characteristics of device A, with a maximum thickness of ~5 nm in the middle of the channel. The as assembled device (dashed curve) shows ambipolar behavior with estimated hole and electron mobilities of 0.21 and 0.06 cm$^2$/Vs respectively. After mild thermal annealing, the mobility value increases to 0.25 and 0.07 cm$^2$/Vs for hole and electron respectively. In addition, the Dirac point shifted from $V_g$ = -20 V to $V_g$=0V upon annealing. Figure 2 (d) shows transfer characteristics of device B with maximum thickness of ~ 14 nm in the middle of the channel. The as-assembled device showed p-type FET behavior with hole mobility of 2 cm$^2$/Vs. Upon mild annealing, it shows ambipolar behavior with Dirac point around zero gate voltage. The estimated hole and electron mobilities are 4 and 1.5 cm$^2$/Vs respectively.

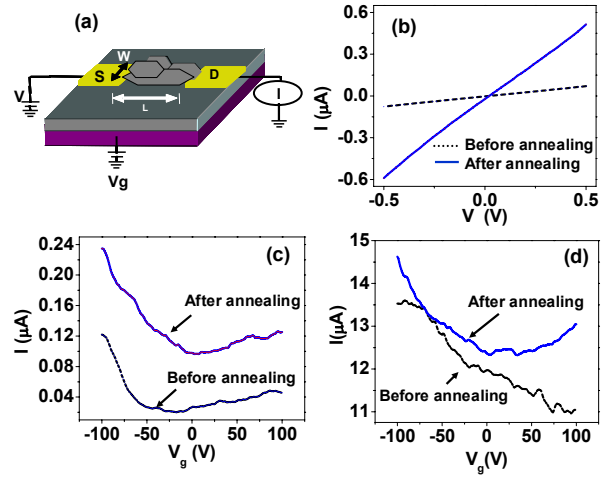

FIG. 2. (Color online) (a) Cartoon of electronic transport measurement set-up. (b) Current-voltage characteristics of a representative DEP assembled sample before and after thermal annealing at 200 $^0$C in Ar/H$_2$ gas. The two terminal resistance improved by almost an order of magnitude from 10 MΩ to 980 kΩ. (c) I-V$_g$ characteristic of one of the RGO FET's before and after thermal annealing. Upon annealing, the mobility improved and dirac point shifted from $V_g$ = - 20 V to 0 V. (d) I-V$_g$ characteristic of another device. This device changed from p-type to ambipolar upon thermal annealing.

Overall, we observed that the thinner devices show ambipolar behavior whereas the thicker ones show p-type behavior before thermal annealing. The p-type behavior in thicker devices before annealing is most likely due to the polarization of trapped water and oxygen molecules between the layers which were removed upon mild annealing, transforming the device to ambipolar [21]. In order to confirm this effect of moisture, a control experiment was performed on one of the devices which showed p-type behavior before annealing and ambipolar after annealing. After subsequent exposure to air for one month, the device transformed from ambipolar to p-type again (see supporting information for evolution of p-type to ambipolar to p-type behavior) [17]. However, the resistance did not go back to the same level recorded before annealing. This result show that the resistance and on-current improvement is not correlated to moisture. So, what causes the resistance improvement and increase of on-current upon mild annealing? Is it due to the improvement of contact resistance by improving mechanical coupling between RGO sheets and electrode, and/or further thermal reduction of already chemically reduced RGO sheets [19]. In order to check whether there was further reduction due to mild annealing, we took XPS data on the RGO films before and after annealing. Figure 3 (a) shows the XPS of the RGO film before annealing while Figure 3 (b) shows the XPS data after annealing. The only peak that shows a change is the C-O bond (286.6 eV, green curve), where we observed the C-C to C-O ratio changed from ~8.0 to 14 before and after annealing respectively. This is an indication that there may be a slightly further reduction due to annealing. Additionally, contact resistance improvement cannot be ignored. We note that contact resistance improvement upon annealing in similar conditions has been observed in DEP assembled carbon nanotube devices, and it is not surprising that DEP assembled graphene devices can also show similar improvement. Additional experiments are needed to separate the effect of reduction and contact resistance. We also took AFM image after the annealing and did not notice any morphology or height change.

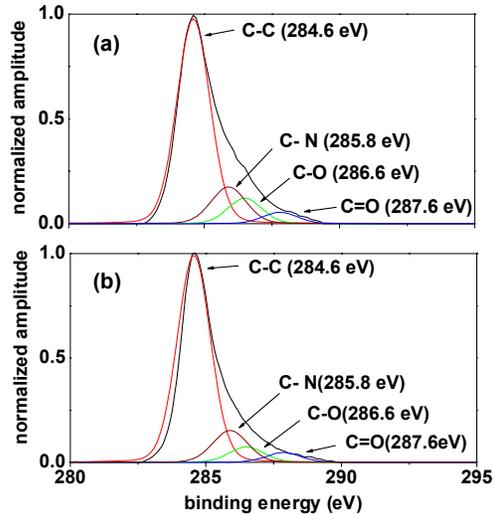

FIG. 3. (Color online) (a) XPS data for RGO film before annealing. (b) XPS analysis of RGO sheets upon annealing at 200 $^o$C in Ar/H$_2$ environment for 1hr. The C-C to C-O ratio is reduced by ~ 40% due to the annealing process.

Overall, 60% of our DEP assembled RGO transistors showed p-type, while 40% of the transistors showed ambipolar behavior before thermal annealing. After mild thermal annealing, all ambipolar RGO FET remained ambipolar with increased hole and electron mobility while 60% of the p-type RGO devices were transformed to ambipolar.

In conclusion, we demonstrate high yield fabrication of field effect transistors (FET) using chemically reduced graphene oxide (RGO) sheets. The RGO sheets suspended in water were assembled between prefabricated gold source and drain electrodes using ac dielectrophoresis (DEP). The two terminal resistances of the devices were improved by an order of magnitude upon mild annealing at 200 $^0$C in Ar/H$_2$ environment for 1 hour. All of the devices showed FET behavior with the application of a gate bias with the majority of them demonstrating ambipolar behavior. The maximum hole and electron mobility of the RGO FET were 4.0 and 1.5 cm$^2$/Vs respectively. High yield fabrication of FETs using graphene nanostructures is a significant step forward in realizing scaled up fabrication of graphene based nanoelectronic devices.

**Acknowledgment.** We thank Paul Stokes for useful discussion. This work has been partially supported by US NSF under grants ECCS 0748091 to SIK and DMR 0746499 to LZ.

# Supplementary Information:
# High yield fabrication of chemically reduced graphene oxide field effect transistors by dielectrophoresis

Daeha Joung[1,2], A. Chunder[1,3], Lei Zhai[1,3], and Saiful I. Khondaker[1,2] *

[1] Nanoscience Technology Center, [2] Department of Physics, [3] Department of Chemistry, University of Central Florida, Orlando, Florida 32826, USA.

**X-ray photoelectron spectroscopy (XPS) spectra of GO and RGO**

We verified the reduction of GO to RGO though XPS. Figure S1 (a) shows the XPS data taken on a thin film of GO dispersed on Si/SiO$_2$. Here we can see the four components of carbon based atom in different functional groups of GO; (i) the non-oxygenated C-C bond (284.5 eV, red curve), (ii) C-O bond (286.6 eV, green curve), (iii) the C=O bond (287.6 eV, blue curve), (iv) C(O)O (289.1 eV, cyan curve). Figure S1 (b) shows the XPS data for a RGO film. GO sheets were chemically reduced in the solution using hydrazine. It can be seen that the peaks for oxygen functional groups were significantly reduced (green and blue curves) and C(O)O bond (cyan curve) that was present in Figure S1 (a) has disappeared in RGO film. An additional peak appears after the reduction at 285.8 eV corresponding to the C in the C-N bond. This also implies that the oxygen in the GO is considerably removed by the reduction process and that nitrogen is now present due to hydrazine treatment [22]. These observations are strong indication of effective reduction of GO to RGO.

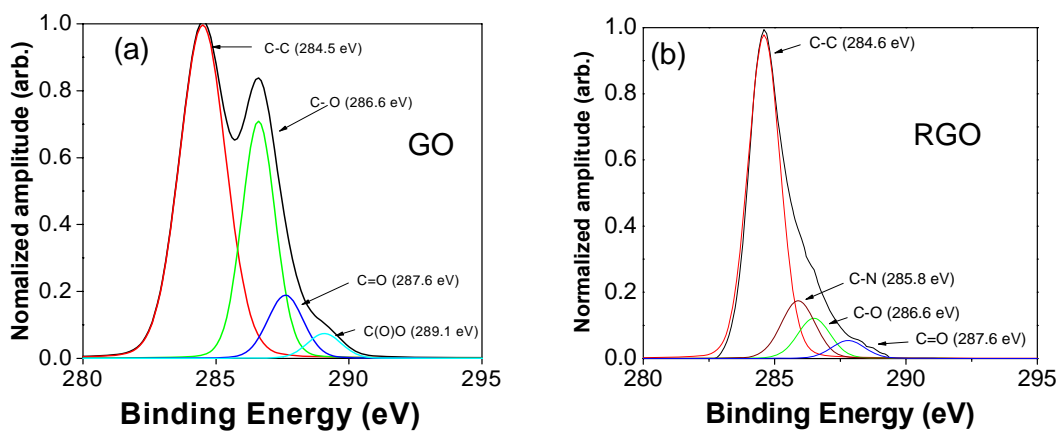

**FIG. S1**. (a) XPS data of a GO thin film on Si/SiO$_2$ substrate. (b) XPS data of a RGO thin film on Si/SiO$_2$ substrate.

**Scanning electron microscope (SEM) images of 100% yield RGO devices**

For the DEP assembly, each of our chips contains 18 electrode pairs (E1-E18). After a small drop of the RGO solution was applied to the chip, an AC voltage of approximately 3 V$_{P-P}$ at 1 MHz was applied with a function generator for 20-30 seconds to the first electrode pair (E1) and then moved to the next pairs. Figure S2 shows SEM images for all 18 electrodes for a representative chip after the assembly. It is clearly seen that the DEP assembly of RGO at each electrode pair is successful, demonstrating 100% yield of our assembly method.

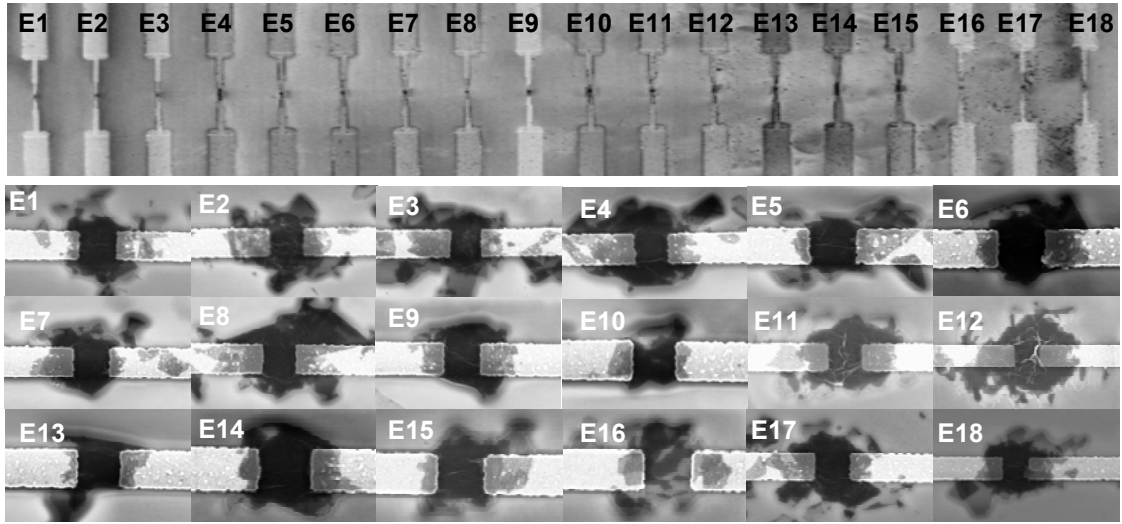

**FIG. S2.** Demonstration of high yield assembly: Top: Scanning electron micrograph (SEM) of an active area of a chip containing after DEP assembly of the RGO sheets. The chip contained 18 pairs of electrodes numbered as E1 to E18. The gap size between each electrode pairs is 500 nm. Bottom: Zoomed in SEM images of all 18 electrode pairs for a clearer picture.

### Evolution of p-type to ambipolar to p-type behavior in RGO-FETs

In order to check the effect of moisture on the transport properties, we carried out a controlled experiment on one of the devices. Figure S3 shows the I-$V_g$ curves for the (i) as-assembled device (blue curve) (ii) after mild thermal annealed device (black dot curve) (iii) exposed in air for 1 month (red curve). The as-assembled device showed p-type behavior initially, while after annealing the characteristic were transformed to ambipolar behavior. When exposed to air for a month, the device recovered its original p-type behavior which indicates that moisture and oxygen are the main sources of the p-type behavior.

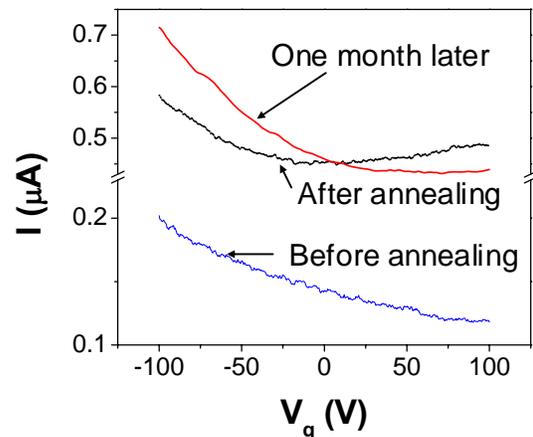

**FIG. S3.** I-$V_g$ characteristics of a representative RGO FET before annealing (blue curve), after mild thermal anneal (black dotted curve), and after subsequent exposure to air for a month (red curve). The device recovered p-type behavior once again after exposure to air.